\documentclass{emulateapj}

\begin{document}

\title{A photodissociation model for the morphology of \mbox{H{\sc i}}\ near OB associations in M33}

\author{Jonathan S. Heiner\altaffilmark{1,2}, Ronald J. Allen\altaffilmark{1}, Pieter C. van der Kruit\altaffilmark{2}
}
\altaffiltext{1}{Space Telescope Science Institute, Baltimore, MD 21218, USA}
\altaffiltext{2}{Kapteyn Astronomical Institute, University of Groningen, PO Box 800, 9700 AV Groningen, the Netherlands}
\email{heiner@astro.rug.nl}


\begin{abstract}
{}
{We present an approach for analysing the morphology and physical properties of \mbox{H{\sc i}}\ features near giant OB associations in M33, in the context of a model whereby the \mbox{H{\sc i}}\ excess arises from photodissociation of the molecular gas in remnants of the parent Giant Molecular Clouds (GMCs). Examples are presented here in the environs of NGC604 and CPSDPZ204, two prominent \mbox{H{\sc ii}}\ regions in M33. These are the first results of a detailed analysis of the environs of a large number of OB associations in that galaxy. We present evidence for ``diffusion'' of the far-UV radiation from the OB association through a clumpy remnant GMC, and show further that enhanced CO(1-0) emission appears preferentially associated with GMCs of higher volume density.}
{}
\end{abstract}


\maketitle

\section{Introduction}

It is widely accepted that stars form from molecular gas in galaxies, and that the largest stars form in Giant Molecular Clouds (GMCs). Such massive stars produce prodigious quantities of far-UV (FUV) radiation, that will react back on the parent GMCs producing photodissociation regions (PDRs) on the cloud surfaces. Highly luminous PDRs very near to OB associations are a rich source of atomic, molecular, and solid grain spectral features, especially in the IR, and several prominent cases (e.g.\ the Orion region) in the Galaxy have been extensively studied both theoretically and observationally \citep[see e.g. the review by][]{1999RvMP...71..173H}. Less well appreciated is that \mbox{H{\sc i}}\ atoms are also produced in such PDRs in quantities sufficient to be detectable in 21-cm radio emission 
(see \citet{2004ApJ...608..314A} and references cited there), and that this can occur in situations of much more modest volume densities and far-UV fluxes than is usually required to produce detectable IR line emission. Furthermore, such relatively low-density PDRs are physically large, rendering them more easily detectable by radio astronomy.

In this paper, we present the first results of a study of \mbox{H{\sc i}}\ features in the the environs of OB associations in M33 (NGC 598) in terms of this PDR model. We explain the observations in a consistent, simple physical context, using a method motivated by the initial discoveries of \citet{1986Natur.319..296A}. Those authors presented evidence that large-scale spiral arm features in the \mbox{H{\sc i}}\ distribution of M83 (NGC 5236), offset from the dust lanes, were produced by photodissociation of \mbox{H$_2$}. \citet{1997ApJ...487..171A} provided a qualitative analysis of similar features in M81 (NGC 3031), consistent with the occurrence of large-scale, low-density \mbox{H{\sc i}}\ PDRs connected to nearby bright FUV sources. They also suggested a more quantitative approach that was subsequently used on M101 (NGC 5457) by \citet{2000ApJ...538..608S}, based on a simple equilibrium model for the physics of \mbox{H{\sc i}}\ production in PDRs. This provided a new method for determining densities of molecular gas in star-forming GMCs of galaxies, a method that is independent of the excitation conditions for specific molecular tracers. Using a combination of GALEX FUV and VLA \mbox{H{\sc i}}\ data, essentially the same technique was applied to M81 \citep{2008ApJ...673..798H} and M83 \citep{2008A&A...489..533H}. While evidence for the presence of widespread molecular gas in both galaxies was found, we concluded that the molecular gas in the GMCs in M81 is generally of lower volume density. This is in agreement with the fact that the CO emission is generally fainter in M81 and hence likely to be less excited \citep[e.g.][]{2006A&A...455..897K}.

The closer proximity of M33 allows for a much more detailed analysis than was possible for M81 and M83, and we are therefore carrying out an extensive study of the \mbox{H{\sc i}}\ in the immediate surroundings of OB associations in this galaxy in the context of the photodissociation model for \mbox{H{\sc i}}\ production. As an example of what can be accomplished with this approach, we analyse the \mbox{H{\sc i}}\ near two prominent OB associations in M33: CPSDP Z204 \citep{1987A&A...174...28C} and NGC 604. The \mbox{H{\sc i}}\ features surrounding these associations are sufficiently bright in 21-cm emission that they stand out above the general background, presenting a plausible case that their morphology is indeed a consequence of dissociation by FUV photons from the cluster of bright young stars nearby. In the case of CPSDP Z204 we model the \mbox{H{\sc i}}\ as a function of depth into a nearby GMC in terms of the ``diffusion'' of the far-UV radiation propagating into a clumpy medium. In all cases we determine gas densities in the surrounding GMCs and compare the results to the appearance of CO(1-0) emission nearby where available. CPSDP Z204 and NGC 604 have in common that local metallicity measurements are available from the literature (which improves the precision of our calculations), as well as CO detections (confirming the presence of molecular gas). We find that the higher density clouds as determined by our method are more likely to show CO emission nearby, in agreement with our earlier M83 results.

Our method and data are described in \S \ref{sec:method}. The NGC 604 and CPSDP Z204 findings are presented in \S \ref{sec:results} and summarized in \S \ref{sec:conclusions}.

\section{Method and Data}
\label{sec:method} 

We have used a simple equilibrium model of photodissociation physics to derive total hydrogen volume densities in candidate PDRs in M33, as we did in \citet{2008ApJ...673..798H} and \citet{2008A&A...489..533H}. Every \mbox{H{\sc i}}\ ``patch'' in the immediate environment of the OB association is considered a candidate PDR. The data entering into the calculation are: the measured far-UV flux of the OB association; the distance from the association to the \mbox{H{\sc i}}\ patch; the column density of the \mbox{H{\sc i}}\ patch; the local dust-to-gas ratio; and, some simplifying assumptions about the actual 3D geometry of the region.

The following basic parameters of M33 were used: We assume a distance of 847 kpc \citep{2006ApJS..165..108S}. A UV foreground Galactic extinction correction of 0.33 was applied, appropriate for the GALEX FUV band, after \citet{1998ApJ...500..525S} and \citet{2007ApJS..173..185G}.

The photodissociation-reformation equilibrium balance is highly sensitive to the local dust content, as it influences both the effective absorption cross section of the dust grains and the \mbox{H$_2$} formation rate. The dust-to-gas ratio $\delta/\delta_0$\ is normalized to the solar neighborhood value. We derive its value directly from the metallicity $12 + \log(O/H)$ as measured in the ionized gas of the OB associations by \citet{2008ApJ...675.1213R}, after \citet{1990A&A...236..237I}. It is assumed that this metallicity measurement from the \mbox{H{\sc ii}}\ region is an acceptable approximation of the metallicity in the molecular clouds that we are interested in.
This is identical to our approach in \citet{2008ApJ...673..798H} and \citet{2008A&A...489..533H}, but now we can take advantage of individual metallicity measurements (where available) instead of using a metallicity gradient over the galaxy. \citet{2008ApJ...675.1213R} point out that, while their gradient fit has a high statistical accuracy, there is significant intrinsic scatter around this gradient (their Fig. 3). We will therefore assume that an individual measurement is a better estimate than using a gradient in this case.

If we adopt a solar metallicity of 8.69 \citep[from][]{2001ApJ...556L..63A}, the dust-to-gas ratio becomes:
\begin{equation}
  \log{\delta/\delta_0} = (12+\log{O/H})_{\mbox{H{\sc ii}}} - (12+\log{O/H})_{\odot}.
  \label{eqn:dtg}
\end{equation}
This is the same as assuming that the metallicity gradient is identical to the gradient in the dust-to-gas ratio, where the gradient is constructed from the solar metallicity (the dust-to-gas ratio scaled to the solar neighborhood value $\delta/\delta_0$ equals 1 by definition) and the value of the metallicity at the location that we are interested in. However, since we are only interested in individual values, a radial dependence is not explicitely computed. We adopted NGC 604's metallicity from \citet{2002ApJ...581..241E} and CPSDP Z204's metallicity from \citet{2008ApJ...675.1213R}.  


The PDR model we applied comes from \citet{2004ASSL..319..731A} with improved coefficients provided by \citet{2004ApJ...608..314A}:
\begin{equation}
  N_{\mbox{H{\sc i}}} = \frac{7.8 \times 10^{20}}{\delta/\delta_0} \ln\left[1+\frac{106G_0}{n}\left(\frac{\delta}{\delta_0}\right)^{-1/2}\right]~\rm{cm^{-2}},
  \label{eqn:n}
\end{equation} 
where $N_{\mbox{H{\sc i}}}$ is the (background subtracted) atomic hydrogen column density (in $\rm{cm^{-2}}$), $G_0$\ is the incident FUV flux measured at the affected \mbox{H{\sc i}}\ patch and
$n = n(\mbox{H{\sc i}}) + 2 n(\mbox{H$_2$})\ \rm{atoms\ cm^{-3}}$
is the total hydrogen volume density. This gas is likely to be mostly atomic on the surface of the GMC and mostly molecular deep inside the cloud. \citet{2004ApJ...608..314A} (their Appendix B) provide a full treatment of how $G_0$\ is defined and derived. \citet{2009ApJ...694..978H} recently proposed an improvement to this model by taking into account a more realistic self-shielding function of molecular hydrogen based upon multiple transitions instead of a single one. In that case, our computed volume densities need to be multiplied by a factor $\left(\delta/\delta_0\right)^{0.2}$, or 0.9 in the case of NGC 604 and 0.8 in the case of CPSDP Z204, well within our estimated level of uncertainty.

We used the publicly accessible GALEX FUV image of M33, which has a linear resolution of about 16 pc ($4\arcsec$). This data at $\lambda \approx 150~\rm{nm}$ is used as a proxy for the dissociating UV ($\lambda \approx 100~\rm{nm}$) since the spectrum of the UV radiation in the ISM is fairly flat, at least in the Galaxy close to the sun \citep{1988ApJ...334..771V}. The 21-cm \mbox{H{\sc i}}\ radio image of M33 was obtained from David Thilker (2007, private communication) and has a linear resolution of 20 pc ($5\arcsec$). Noise levels are typically $2 \times 10^{20}\ \rm{cm}^{-2}$.
At these spatial scales we expect to start seeing more frequently the signature PDR morphology in the \mbox{H{\sc i}}\ features found near FUV sources. The radio data were taken with the VLA and with the GBT \citep[combined using a similar procedure as described in][]{2009ApJ...695..937B}. They are therefore photometrically accurate (within the noise levels). We adopt a local \mbox{H{\sc i}}\ background level for each region, based on measurements of the \mbox{H{\sc i}}\ column densities away from the candidate PDRs. The background levels are measured on a scale of typically 100 pc, and can be as high as $1 \times 10^{21}\ \rm{cm^{-2}}$. The remaining column density after subtraction of the background is then assumed to be produced entirely by photodissociating photons from the nearby OB association\footnote{The origin of the smooth \mbox{H{\sc i}}\ background is not discussed here. It is possible that this too is photodissociated gas, arising for example from an extended population of B stars, but some fraction of it may be primordial.}.

To apply the PDR model, we have generally followed the methods previously used in M81 and M83. However, owing to the higher linear resolution, some further improvements have been made. The UV sources at the center of each region are identified (to the extent that they are resolved), their fluxes measured, and used to calculate a cumulative incident flux $G_0$ on the \mbox{H{\sc i}}\ patch of current interest. No internal extinction correction was applied, for reasons outlined in \citet{2008ApJ...673..798H}. We measured the distances between the central UV sources and a number of nearby \mbox{H{\sc i}}\ 'patches', defined as a local maximum in the \mbox{H{\sc i}}\ column density. These patches were generally selected based on their prominence in the image and/or their proximity to available CO detections. In these measurements we limited the distance between any UV source and any \mbox{H{\sc i}}\ patch to 200 pc, which is smaller than the cut-off of 400 pc we used previously, but more appropriate to the scale of the candidate PDRs we expect to see at this level of detail. At these separations, the FUV flux contributed by the young clusters of OB stars has also dropped to about 1\% of the local ambient UV flux, and the additional \mbox{H{\sc i}}\ contributed by those clusters will be difficult to distinguish in the noise and confusion of the general \mbox{H{\sc i}}\ background. The final result of the calculation on each \mbox{H{\sc i}}\ patch is an estimate of the total hydrogen volume density $n = n(\mbox{H{\sc i}}) + 2 n(\mbox{H$_2$})\ \rm{atoms\ cm^{-3}}$ in the GMC associated with that patch.

\section{Results}
\label{sec:results} 

\subsection{Atomic hydrogen in candidate PDRs}

\begin{figure}[bt]
  \centering
  \includegraphics[width=0.4\textwidth]{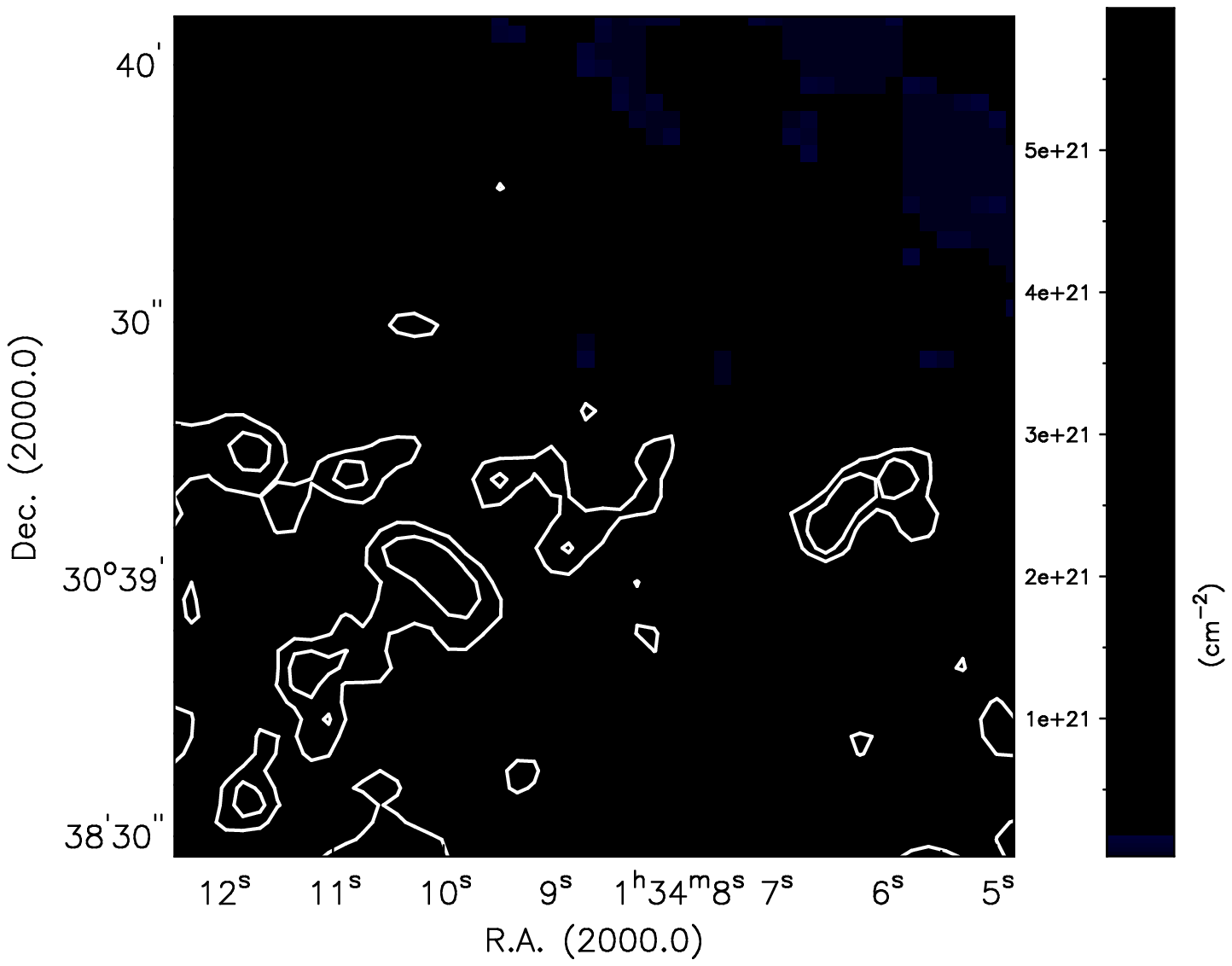}
  \includegraphics[width=0.4\textwidth]{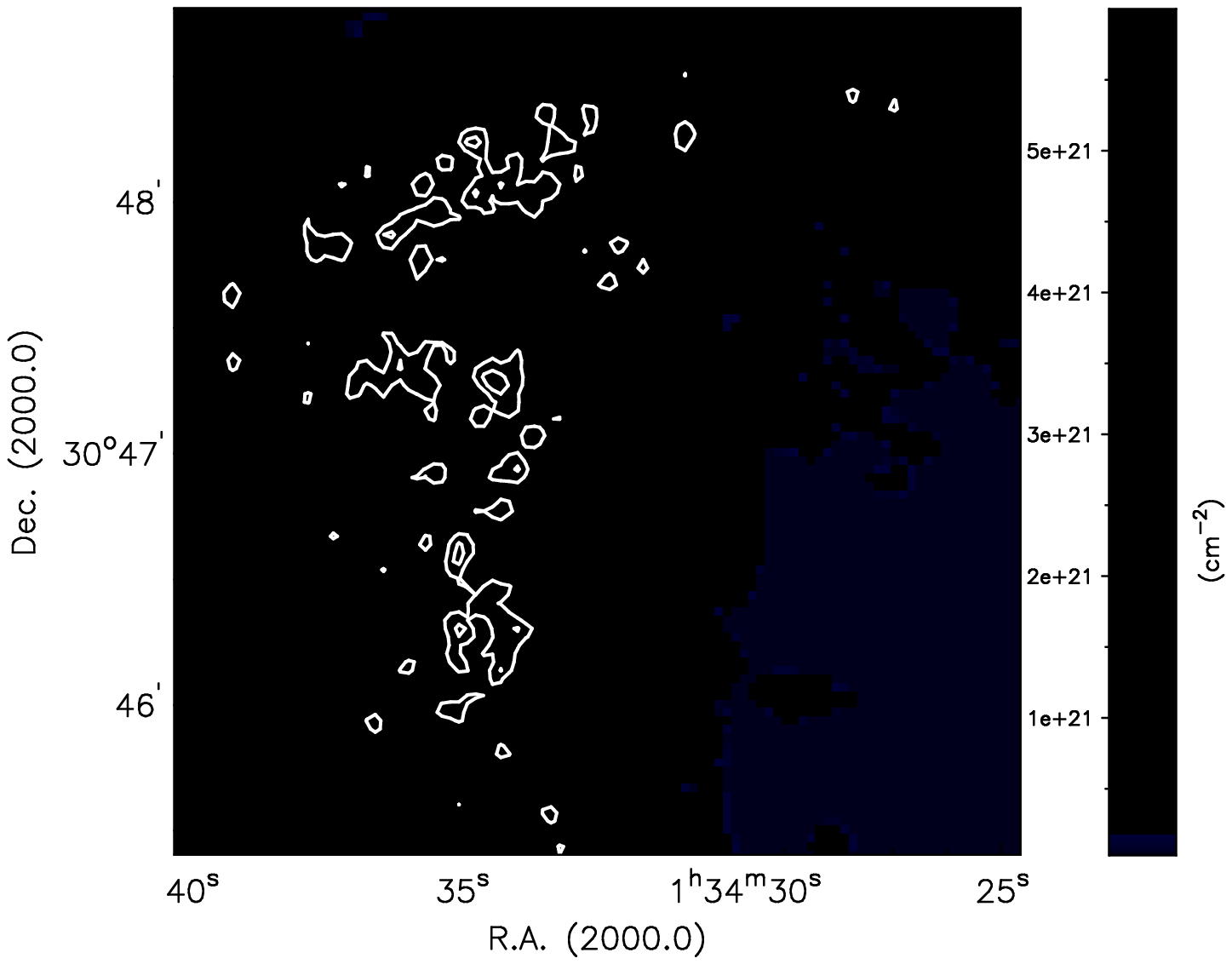}
  \caption{\label{fig:HImaps} Atomic hydrogen maps with selected contours. CPSDP Z204 (top) and NGC 604 (bottom) are shown. The central location of the OB associations are marked with a black cross.}
\end{figure}

As a prelude to the actual analysis of the candidate PDRs, we first present the atomic hydrogen images of the immediate surroundings of the two OB associations that we selected. The general morphology of the \mbox{H{\sc i}}\ will, in our picture, be a function of both the local surface number density of GMCs and the ambient UV flux. Since we are assuming that atomic hydrogen peaks near OB associations are produced by photodissociation, we can measure these peaks and then use these results to calculate total hydrogen volume densities. It should be noted that we do not aim to identify all \mbox{H{\sc i}}\ patches, but rather attempt to find those that are clearly produced by photodissociation.

The CPSDP Z204 area shows a variety of \mbox{H{\sc i}}\ patches with a suggestive shell-like feature immediately surrounding the OB association. Two more extended patches can be seen further away, as well as some more patches that either do not stand out much or are beyond our 200 pc cut-off. The more extended patches are studied in more detail in this paper, because they are likely to be caused by the central OB association as well as being sufficiently resolved to try to model the \mbox{H{\sc i}}\ distribution radially.

The distribution of atomic hydrogen in NGC 604 is part of a larger spiral arm and appears to curve around the central OB association in NGC 604 on the eastern side. While the general level of \mbox{H{\sc i}}\ column density is fairly high, the peaks do show a bit of structure, albeit a chaotic one. We analyzed four \mbox{H{\sc i}}\ patches in this area, as well as considering a larger-scale approach to the PDR morphology by averaging the atomic hydrogen over a larger area. These patches were chosen partially because of the presence of CO in the area (a confirmation of the presence of molecular gas, sure to produce atomic hydrogen under the influence of FUV emission). Another reason to look at NGC 604 in particular is the wealth of available data. Since our method is generally sensitive to low-density molecular gas, it is interesting to test it at the extreme end of higher densities that are known to occur in NGC 604 and compare the results.

\subsection{\mbox{H{\sc i}} morphology in the environs of CPSDP Z204}
\label{sec:Z204}

\begin{figure*}[t]
  \centering
  \includegraphics[width=0.4\textwidth]{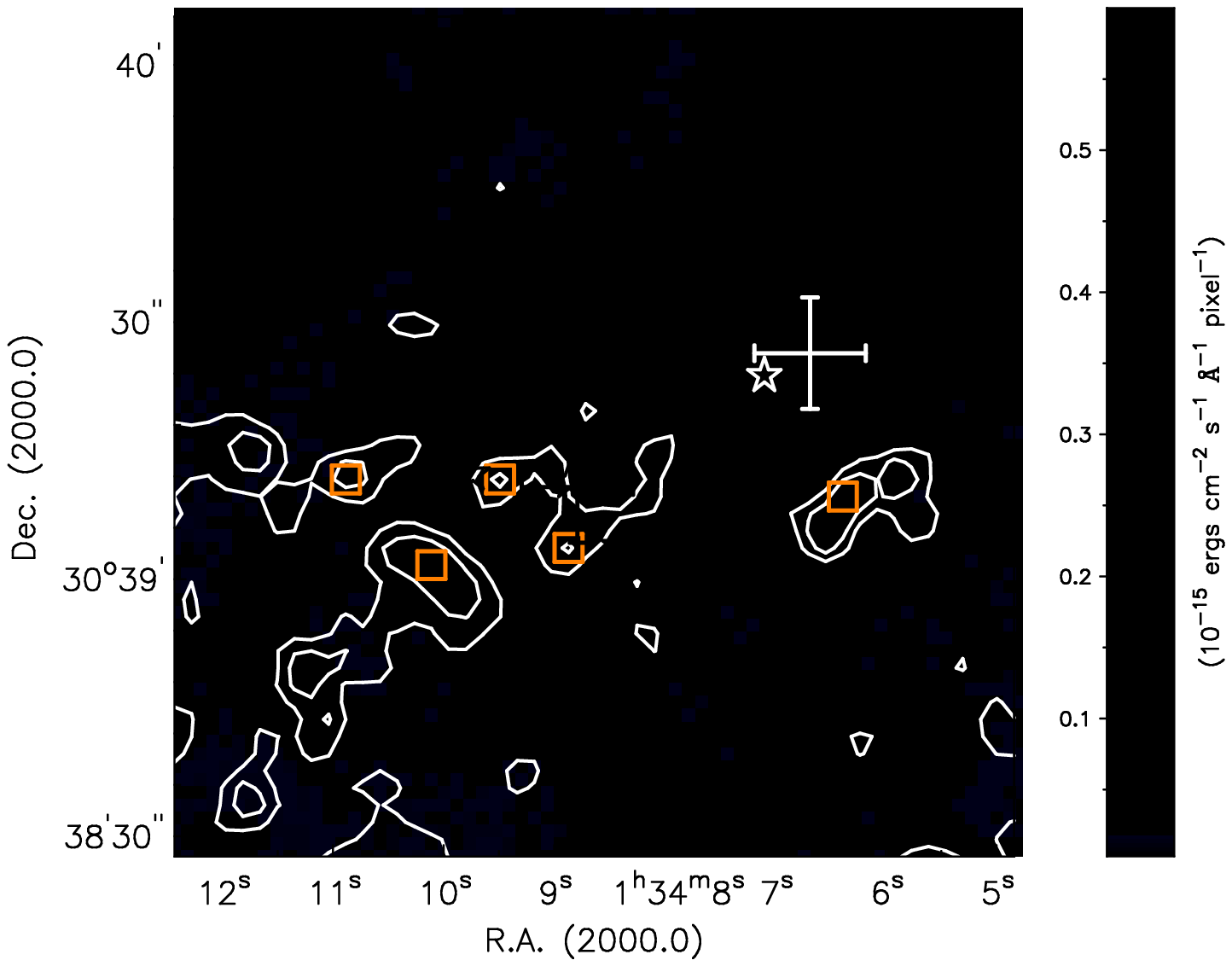}
  \includegraphics[width=0.332\textwidth]{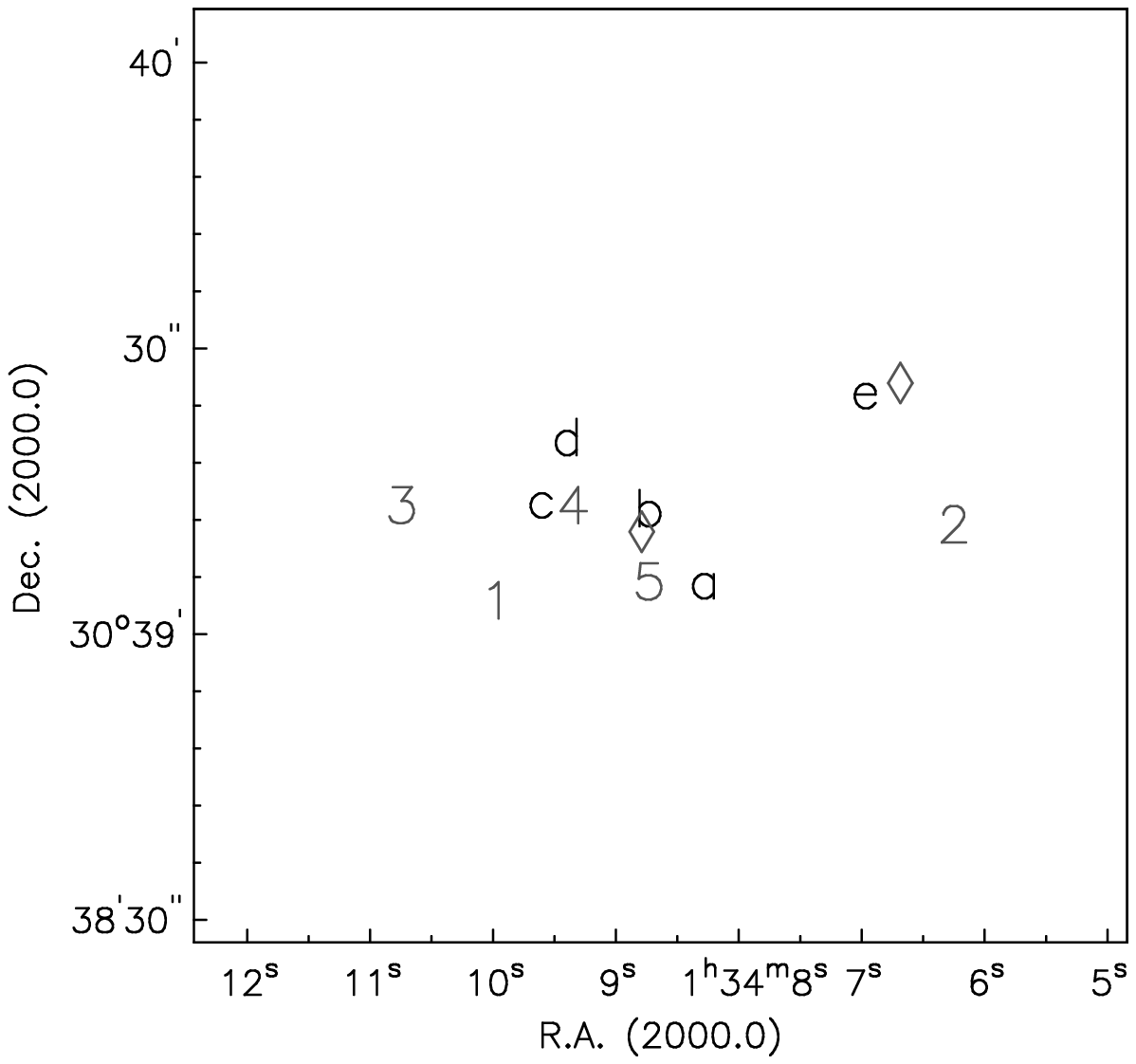}
  \caption{\label{fig:cpsdpz204} Left panel: CPSDP Z204 at full resolution. The FUV color image is overlaid with a selection of \mbox{H{\sc i}}\ contours. These contours (not background-subtracted) are at 3 and 4 $\times 10^{21}~\rm{cm}^{-2}$ only, and were chosen to emphasize the morphology of the \mbox{H{\sc i}}\ peaks. The noise level is about 2 $\times 10^{20}~\rm{cm}^{-2}$. Stars mark the locations where the FUV fluxes were measured. The orange boxes are the locations of the measured \mbox{H{\sc i}}\ patches. Two crosses mark the locations of CO detections by \citet{2003ApJS..149..343E}, at a resolution of 13\arcsec\ (50 pc). Right panel: Finding chart of the same region with letters (UV sources) and numbers (\mbox{H{\sc i}} patches) corresponding to the results in Table \ref{tab:results}. The CO detections are the crosses near sources b and e.}
\end{figure*}

The \mbox{H{\sc ii}}\ region CPSDP Z204 (galactocentric radius 1.63 kpc) was included in several surveys---most notably the one from which it derives its name \citep{1987A&A...174...28C}, but it does not appear to have been singled out for individual study before. 

We have modeled the morphology of two particular \mbox{H{\sc i}}\ patches near FUV sources using PDR physics and simple models of the geometry of the GMCs. We identified four FUV sources which define the center of Z204, and one additional source a bit further away to the northwest as shown in Fig. \ref{fig:cpsdpz204}. We also located five \mbox{H{\sc i}}\ local maxima, that we assume to be on the surface of GMCs. The FUV fluxes are generally about ten times fainter than in NGC 604. CO emission detected by \citet{2003ApJS..149..343E} can be seen in two locations, at a resolution of 13\arcsec\ or about 50 pc. The results of our measurements of CPSDP Z204 are listed in Table \ref{tab:results}. The letters and numbers in the finding chart correspond with the superscripts of values in the table. The resulting total hydrogen volume densities range from 15 to $295\ \rm{cm}^{-3}$. One of the CO detections, close to source b, is accompanied by two \mbox{H{\sc i}}\ patches (nos. 4 and 5) that yield the highest volume densities in the area. The other CO detection, close to source e, does not correspond to a particular \mbox{H{\sc i}}\ patch at the highest column densities, but its presence is a confirmation that there is molecular gas in the area. This is important for the applicability of our method. It should also be noted here that the resolution of the CO measurements is poorer than that of the \mbox{H{\sc i}}\ map, and the CO location is accurate to within about 80--90 pc at best.

\begin{figure}[hbt]
  \centering
  \includegraphics[width=0.4\textwidth]{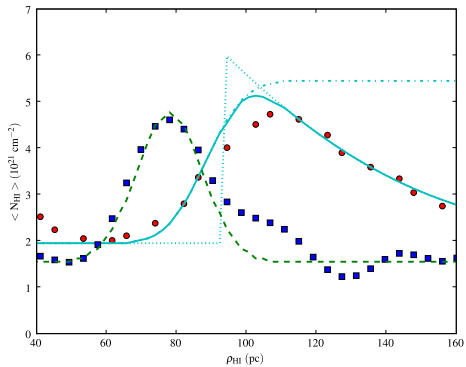}
  \caption{\label{fig:HIpatch} The average \mbox{H{\sc i}}\ column density distribution of CPSDP Z204 patches 1 (circles) and 2 (squares)---see the finding chart in Fig. \ref{fig:cpsdpz204}. The dashed-dotted line represents an infinite edge fit to patch 1 smoothed by the telescope beam. The dotted line is the expected underlying distribution for photodissociated \mbox{H{\sc i}}\ produced on the surfaces of a clumpy GMC as the FUV radiation diffuses into it, and the solid line is the result of smoothing that distribution with the telescope beam. The dashed line is the fit to patch 2, assuming that this \mbox{H{\sc i}}\ patch is $3''$ wide in the direction of this plot and hence only slightly resolved by our $5''$ beam. See text for more details of the models.}
\end{figure}

The \mbox{H{\sc i}}\ patches in the area are consistent with being the photodissociated edges of GMCs, either unresolved or partially resolved, and illuminated by the central cluster of OB stars. To test this assumption we have modeled two patches numbered 1 and 2 in the finding chart of \ref{fig:cpsdpz204}. 

The average \mbox{H{\sc i}}\ column densities of these patches were measured along the line through the center of the patch in the direction to the nearest cluster of UV sources. The  column densities were averaged in thin slabs of $1'' \times 12''$ oriented perpendicular to that line. Distances were deprojected in the case of patch 1 by a factor of 0.6 in order to obtain the separation $\rho_{\mbox{H{\sc i}}}$ in the plane of the galaxy. The resulting average column densities are shown in Fig.\ \ref{fig:HIpatch}. 

The density profile of patch 2 (squares) is the easiest to model. The patch seems unresolved along the line of sight from the UV source, and barely resolved in the perpendicular direction. We fitted the data to a model of a simple rectangular \mbox{H{\sc i}}\ patch of constant column density, convolved with our 5\arcsec\ (radio) beam. The best result is achieved with a 3\arcsec\ $\times$ 9\arcsec\ patch with an intrinsic column density of $7 \times 10^{21}\ \rm{cm^{-2}}$ (not including a background level of $1.6\times 10^{21}\ \rm{cm^{-2}}$). The volume density inferred for the GMC underlying this patch is $15\ \rm{cm^{-3}}$ using the observed column density enhancement of about $3 \times 10^{21}\ \rm{cm^{-2}}$ from Fig.\ \ref{fig:HIpatch}, but this decreases to $4\ \rm{cm^{-3}}$ after correcting the column density for beam smoothing. Note further that the brightness temperature associated with such a level of \mbox{H{\sc i}}\ column density suggests that the \mbox{H{\sc i}}\ is quite possibly optically thick, in which case the real column density may be significantly higher. The extra ``lump'' of \mbox{H{\sc i}}\ gas at about 100 pc distance from the FUV source might be an indication that the GMC is somewhat ``porous'' to the UV radiation, which has penetrated through holes in the GMC and produced more atomic hydrogen further from the FUV source. A more elaborate version of this model will now be described for patch 1.

Patch 1 looks resolved in Fig.\ \ref{fig:HIpatch}, but cannot be explained by a simple rectangular source as we have done for patch 2; the \mbox{H{\sc i}}\ distribution drops off too slowly on the right side, beyond about 100 pc (circles). The approximately exponential decrease at larger distances from the FUV source suggest a model GMC which is porous to FUV radiation, with many unresolved molecular clumps embedded in a lower-density medium \citep[see e.g.][]{1996ApJ...472..191F}. These molecular clumps will all have photodissociated \mbox{H{\sc i}}\ on their surfaces; it is this \mbox{H{\sc i}}\ which we observe. The effective FUV optical depth of such a medium is much lower than that expected for uniform gas. 

We calculate first the FUV flux incident on the GMC at about 90 pc using the FUV luminosity of the OB associations and the distance to the GMC. Beyond that point the FUV flux drops not only as the square of the distance, but also with an extra internal ``extinction'', proportional to $e^{-\kappa r}$. In this case an opacity of $\kappa = 0.024\ \rm{pc}^{-1}$ (optically thin) gives a reasonable fit. According to \citet{1999RvMP...71..173H}, a clumpy medium that is essentially empty in the interclump regions has an effective $\kappa_e = p_0 \kappa_0$. The filling factor is approximately $p_0 = 0.003$ for GMCs that have densities comparable to what we are finding, which means that $\kappa_0$ would be $\approx 8$; this is consistent with the properties of small, dense molecular clumps such as those found in the Orion Nebula. 

The final step in our model is to smooth this source distribution with the $5''$ beam, resulting in the solid line shown in Fig.\ \ref{fig:HIpatch}. We note that this model provides a good fit to the data (circles).

\subsection{NGC 604}
\label{sec:N604}

\begin{figure*}[t]
  \centering
  \includegraphics[width=0.7\textwidth]{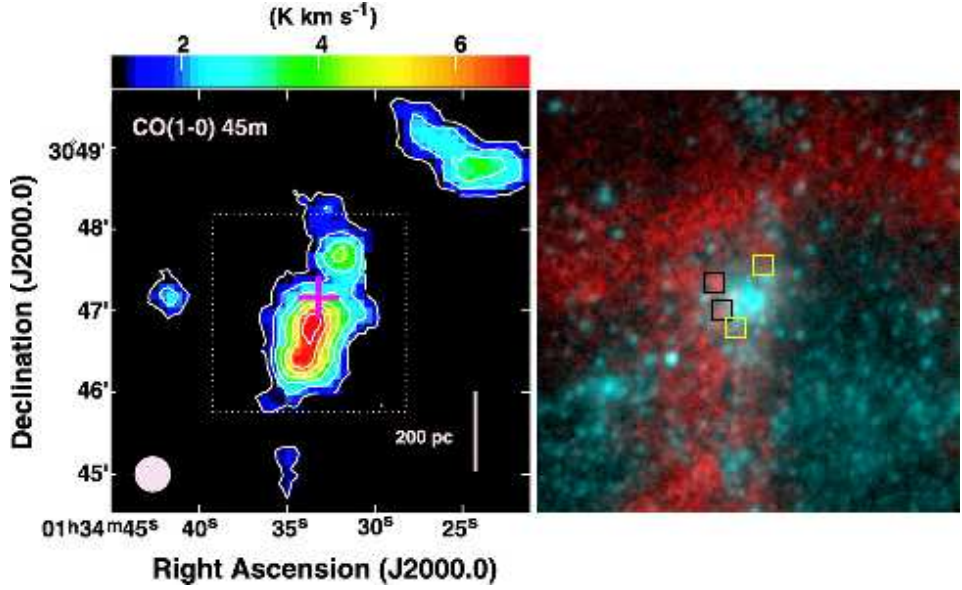}
  \caption{\label{fig:ngc604} Left panel: NGC 604 CO(1-0) image, reproduction from \citet{2007ApJ...664L..27T}. The morphology of the warm dense molecular gas is clearly visible. Their CO(3-2) emission shows a similar morphology, tracing high excitation areas. The cross marks the center of NGC 604. Right panel (same scale): FUV (blue) and \mbox{H{\sc i}}\ (red) qualitative image at arbitrary scale (not background-subtracted), showing the distribution of atomic gas and young stars. The four boxes indicate the locations of select \mbox{H{\sc i}}\ patches. The yellow ones are close to the CO emission peaks.}
\end{figure*}

\begin{figure*}[tb]
  \centering
  \includegraphics[width=0.4\textwidth]{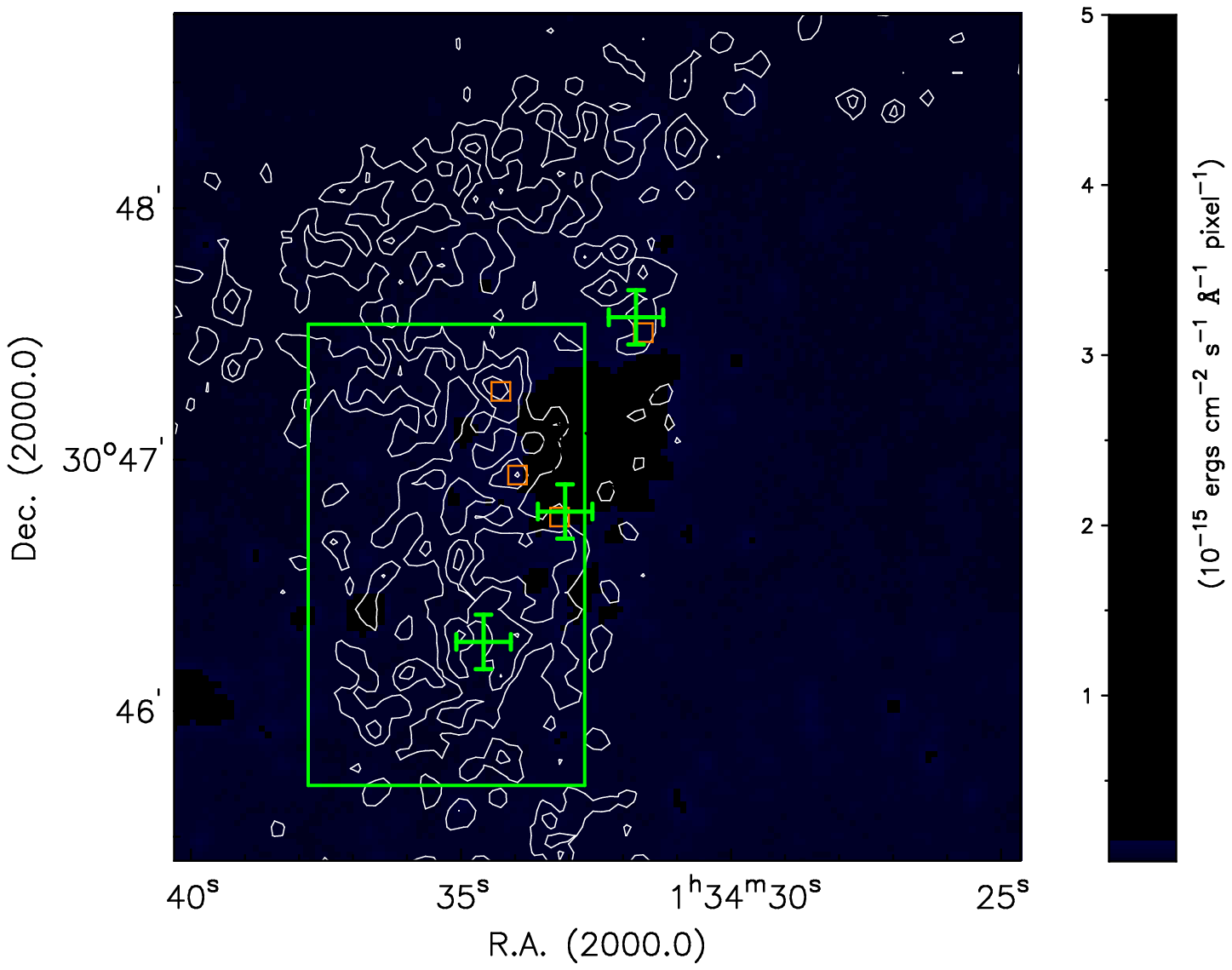}
  \includegraphics[width=0.338\textwidth]{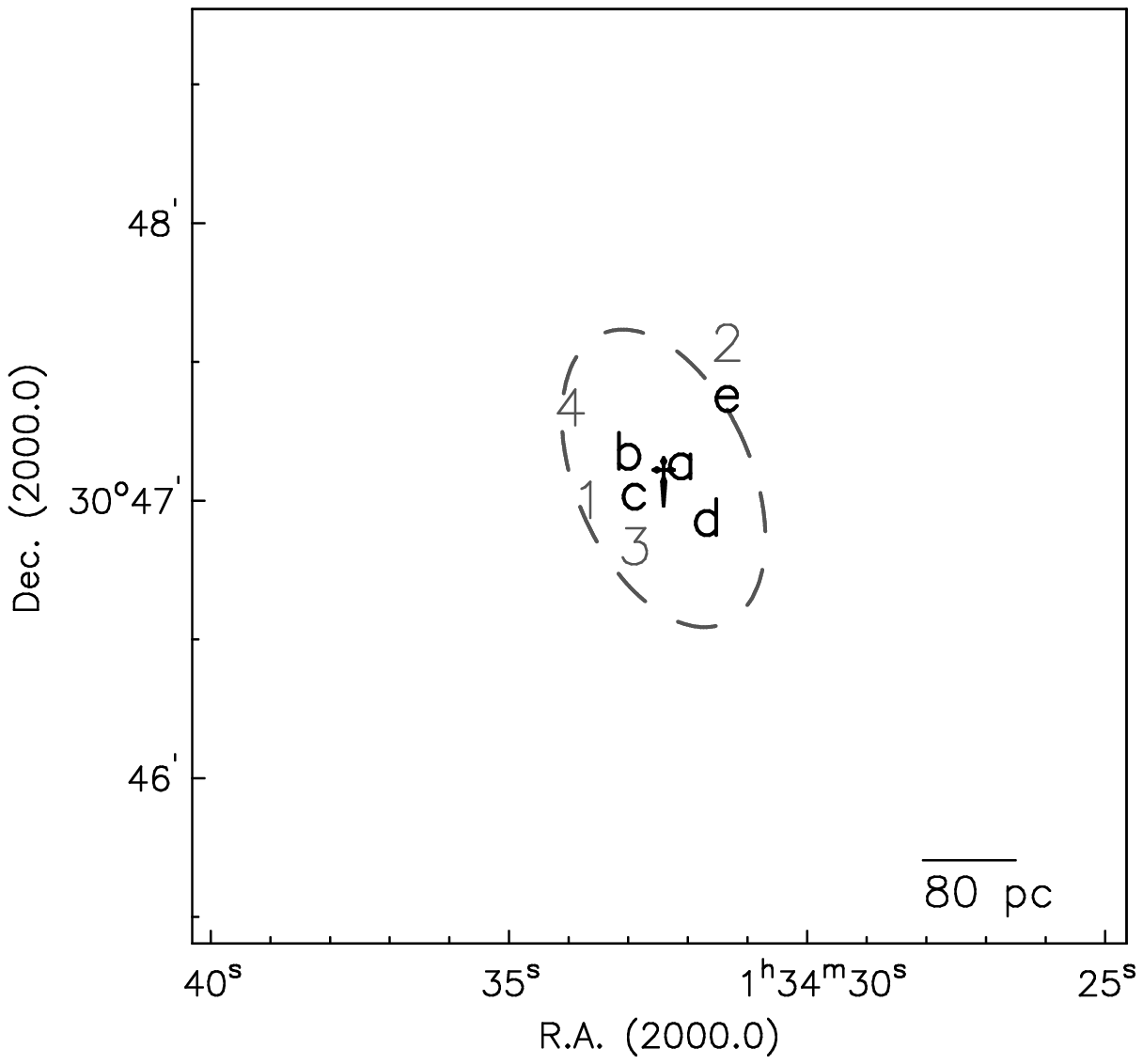}
  \caption{\label{fig:ngc604_2} NGC 604 region at full resolution. The FUV flux range is indicated in the color bar. The \mbox{H{\sc i}}\ contours are 3, 4 and 5 $\times 10^{21}~\rm{cm}^{-2}$. The green rectangle indicates the region where the \mbox{H{\sc i}}\ columns were averaged for our measurement. The orange boxes are the locations of the measured \mbox{H{\sc i}}\ patches. The green crosses mark the presence of CO. See Table \ref{tab:results} for the measurements corresponding to the finding chart.}
\end{figure*}

NGC 604 is the largest and most luminous \mbox{H{\sc ii}}\ region in M33 and a site of active star formation, situated at a galactocentric radius of 3.47 kpc.
It consists of several clusters of O and B stars \citep{1999ApJ...514..188C}, which are partially resolved in the GALEX image of NGC 604. The region is known to contain molecular clouds of various sizes in a clumpy medium \citep{2006A&A...459..161L,1992ApJ...385..512W}.
Finally, \citet{1992ApJ...385..512W} pointed out that M33 \mbox{H{\sc i}}\ data implies that NGC 604 has a relatively large mass fraction of atomic hydrogen, which may be the result of photodissociation. This makes the region an excellent target to find large-scale PDRs. We now estimate molecular gas densities in NGC 604 using our model, and compare our results with those obtained from CO maps.

A detailed view of NGC 604 in FUV and \mbox{H{\sc i}}\ 21-cm radio emission is shown in the right panel of Fig.\ \ref{fig:ngc604}. The FUV emission (blue) is spread across the region, but it is strongest at the \mbox{H{\sc ii}}\ region defining NGC 604. The atomic hydrogen (red) follows one of M33's spiral arms, and appears to ``wrap around'' to the east of NGC 604, forming a large \mbox{H{\sc i}}\ structure of typical size 200 pc similar to those structures first identified in M81 by \citet{1997ApJ...487..171A}. The \mbox{H{\sc i}}\ to the west of the arm drops below the detection limit, but it is present at faint levels in lower resolution maps. Five resolved OB associations were identified and their FUV fluxes measured. Four representative \mbox{H{\sc i}}\ patches were selected as candidate PDRs, two of them based on the detection of CO close to their locations. This is shown in Fig.\ \ref{fig:ngc604_2}. Table \ref{tab:results} lists the measurements of candidate PDRs, taken in a similar fashion as in CPSDP Z204. The resulting total hydrogen densities of the underlying GMCs vary from about 20 to 140 $\rm{cm}^{-3}$, with the highest values corresponding to the two locations near the CO detections by \citet{2003ApJS..149..343E}.  

We have also analyzed the \mbox{H{\sc i}}\ ``envelope'' around NGC 604 as if we had observed it's \mbox{H{\sc i}}\ with a typical resolution of $\sim 100 - 200$ pc, analogous to our earlier work on M101, M81 and M83 (boldface in Table \ref{tab:results}). At this linear resolution, the (smoothed) \mbox{H{\sc i}}\ column densities are lower, leading to higher GMC volume densities of typically 500 $\rm{cm}^{-3}$ estimated with a simple application of the PDR model. However, the more detailed analyses we have provided in this paper show that the true mean densities are likely to be significantly lower when a clumpy source geometry and the effects of finite telescope resolution are included\footnote{This apparently counter-intuitive behavior is a consequence of Equation \ref{eqn:n}, which shows that for a given $G_0$ the only way to increase $N_{\mbox{H{\sc i}}}$ is to decrease the total volume density $n$.}.



The densities of the GMCs that we found here are lower than those obtained using CO methods. This is caused by our method being specifically sensitive to lower densities. Higher density clouds have a thinner layer of atomic hydrogen, as well as a lower beam filling factor as they are more concentrated. The GMCs with the highest densities that can be detected through their CO emission will therefore correspond to smoothed \mbox{H{\sc i}} column densities that are too low to stand out from the general background level. What can be seen here is that our method yields cloud densities on the high end of the range of densities to which we are sensitive.

\begin{table*}[bth]
  \caption{\label{tab:results} Measurements of NGC 604 and CPSDP Z204}
  \centering
  \begin{tabular}{lll}
    \hline\hline
    & NGC 604 & CPSDP Z204\\
    \hline
    $\delta/\delta_0$			& 0.63 
                                        & 0.38 \\
    FUV fluxes ($10^{-15}\ \rm{ergs\ cm^{-2}\ s^{-1}\ \AA^{-1}}$)	& $406^a$, $15.3^b$, $22.6^c$, $3.47^d$, $\mathbf{914}$ 
					& $6.03^a$, $6.73^b$, $9.61^c$, $3.72^d$, $10.9^e$ \\
					& $8.54^e$ & \\
    $N_{bg}\ (10^{21}\ \rm{cm}^{-2})$	& 1.0 
    					& 1.0 \\
    $N_{\mbox{H{\sc i}}}\ (10^{21}\ \rm{cm}^{-2})$	& $4.2^{1:abcd}$, $2.6^{2:ade}$, $3.0^{3:abcd}$, $4.6^{4:abcd}$, $\mathbf{2.0}$
    					& $4.6^{1:abcd}$, $4.4^{2:e}$, $3.6^{3:abcd}$,\\
    \phn\phn(background-subtracted)	& 
					& $3.4^{4:abcde}$, $3.2^{5:abcd}$ \\
     $G_0$ (cumulative)			& $7.40^1$, $6.06^2$, $10.53^3$, $6.93^4$, $\mathbf{15.21}$
					& $0.87^1$, $0.65^2$, $0.45^3$, $7.20^4$, $3.97^5$ \\
     $G/G_{bg}$ range			& $0.01-4.45^1$; $<0.01-4.01^2$; $0.01-6.27^3$;  
     					& $0.02-0.18^1$, $0.57^2$, $0.01-0.09^3$,\\
					& $<0.01-4.45^4$; $\mathbf{6.49}$ 
					& $0.13-1.65^4$, $0.05-0.51^5$ \\
    n (derived, in $\rm{cm}^{-3}$)	& $34^1$, $109^2$, $139^3$, $22^4$, $\mathbf{539}$
					& $18^1$,   $15^2$,   $16^3$,   $295^4$,  $180^5$ \\
    Fractional error range		& 0.26 - 0.44 
					& 0.26 - 0.36 \\
    \hline
    \multicolumn{3}{l}{A description of this table can be found in \S\S \ref{sec:Z204} and \ref{sec:N604}.}
  \end{tabular}
\end{table*}

\section{Conclusions}
\label{sec:conclusions} 

\begin{enumerate}
  \item Using VLA + GBT observations of the 21-cm \mbox{H{\sc i}}\ line and GALEX observations of the FUV emission, along with a simple model for the photodissociation of molecular hydrogen on the surfaces of GMCs, we have estimated total hydrogen volume densities in the GMCs surrounding the two OB associations NGC 604 and CPSDP Z204 in M33.
  \item We find values for the molecular gas densities in the parent GMCs near these OB associations ranging from 15 to 540 $\rm{cm}^{-3}$. Regions with higher gas densities ($\gtrsim 100~\rm{cm}^{-3}$) appear to be correlated with the presence of CO(1-0) emission.
  \item We explain the morphology of two \mbox{H{\sc i}}\ patches in the environs of CPSDP Z204 using straightforward assumptions about the underlying geometry of the molecular gas. One patch appears to be a slightly-resolved GMC. The other source was largely resolved; it can be modeled as a clumpy molecular medium which is very porous, and therefore significantly more transparent to FUV emission than would otherwise be expected. In our model, the \mbox{H{\sc i}}\ we are observing is produced by photodissociation of H$_2$ on the surfaces of these molecular clumps.
\end{enumerate}

Our approach views the \mbox{H{\sc i}}\ found in the environs of star-forming regions of galaxies to be a \textit{product} of the star formation process, not a precursor to it. The ability to model the morphology of \mbox{H{\sc i}}\ features in the environs of OB associations and to predict the appearance of CO emission as an excitation effect related to higher volume densities in the parent GMCs are two results of this paper which lend further support to this approach.
A more extensive treatment of GMCs in M33 based on this approach is in preparation (Heiner 2009, PhD thesis).

\begin{acknowledgements}
JSH acknowledges the support of a Graduate Research Assistantship provided by the STScI Director's Discretionary Research Fund. We are grateful to David Thilker and Robert Braun for making their M33 radio data available to us. David Thilker also kindly provided comments on early versions of this paper.
\end{acknowledgements}


\end{document}